\documentclass[
aps, pre,
twocolumn, 
showpacs, 
nopreprintnumbers, 
amsmath,amssymb,
floatfix,
letterpaper
]{revtex4}

\usepackage{graphicx}
\usepackage{dcolumn}
\usepackage{bm}

\begin{document}


\title{ 
Parameterized excitation operators for coupled cluster method
}

\author{Quanlin \surname{Jie}}
\email[E-mail: ]{qljie@whu.edu.cn}
\affiliation{%
Department of Physics, Wuhan University,
Wuhan 430072, P. R. China
}%

\date{\today}

\begin{abstract}
We present a coupled cluster method (CCM) with optimized excitation operators.
The efficiency comes from a parameterized form of excitation operators. The
parameters are found by variational optimization procedure. The resultant number
of excitation operators is much smaller than that of the conventional CCM
theory.  This property makes it possible to apply the method in systems of solid
state physics. Starting from Hartree-Fock state as the reference state, i.e.,
the Fermi sea, we search for particle-hole excitation operators such that the
wave function of configuration interaction in terms of these excitation
operators spans a good approximation to the ground state. The Match-pursuit
algorithm is capable of doing the search of the excitation operators. The
resultant operators are our excitation operators for the CCM wave function. We
test the method by two dimensional fermionic Hubbard model on a square lattice.  
\end{abstract}

\maketitle

The coupled cluster method (CCM)~\cite{1,1a,1b} is an valuable first principle
numeric tool to treat fermionic many body systems. It is now popular in
Chemistry community, although it was originally proposed by nuclear physicist in
the fifties of last century~\cite{2,3}. For moderate sized atoms and molecules,
CCM is cost effective with remarkable accuracy. Some comments refer CCM as a
best wave function based numeric method. As a post Hartree-Fock theory, the CCM
is an improvement to the mean field result. It uses the determinant state of the
mean field result as reference state (other choices of the reference state are
possible, see, e.g. the multi-references CCM~\cite{4,5,6,7,8,9,10}). Correlation
is taken into account from particle-hole excitations from the Fermi sea, i.e.,
the reference state. A closely related theory with the same idea of
particle-hole excitations from a Fermi sea is the configuration interaction (CI)
theory~\cite{1}. The reference state plus all possible particle-hole excitations
form a complete basis set of the Hilbert space. Taking into account of all
possible excitations, i.e., full configuration interaction (FCI), leads to exact
solution. This is only possible for small systems. Practical applications of CI
usually implement a truncated form of FCI that consider only certain order of
excitations. However, this truncated CI is not size extensive. The accuracy
becomes poor for large systems. CCM, on the other hand, takes into account of all
orders of particle-hole excitations in a concise way by an exponential function
of some low order excitation operators. In other words, the amplitudes of high
order excitations in CCM are constructed from low order ones. As a result, CCM is
size extensive. In practical calculations, considering only single and double
excitations can reach accurate results in many cases. A useful property of CCM
is that the final result of CCM is not quite sensitive to the details of
excitation operators that form the exponent as long as enough excitations are
considered. 

The CCM has a drawback that prevents it from widespread applications in condensed
matter physics. The scaling of CCM with system size is a very steep polynomial.
For example, if only single and double particle-hole excitations are considered,
the scaling is about $N^6$ with $N$ the dimension of single particle Hilbert
space. Thus it is impractical to do calculation with CCM for large systems,
including many important systems of condensed matter physics. However, by
exploiting the translational symmetry, some special cases, such as Hubbard model
at half filling~\cite{11,12,13} and various spin systems modeled by Heisenberg
model~\cite{14,15,16,17,18} can  be treated by CCM.  There are many active
investigations to extend the applicability of
CCM~\cite{19,20,21,22,23,24,25,26,27,28}. The basic idea is to restrict the
selection of excitation operators. Up to now, almost all methods employ
excitation operators that are orthogonal with each other, i.e., the states
resultant from excitation operators acting to the reference state are orthogonal
with each others.

Here we explore a new approach to the CCM by employing an efficient form of
particle-hole excitation operators. These excitation operators for the exponent
of CCM wave function are non-orthogonal with each other. All possible states
resultant from such excitation operators acting on the reference state form a
over complete basis set. The Matching-pursuit algorithm~\cite{29,30,31,32} is
capable of searching this over complete basis set to find most relevant
excitation operators for the coupled cluster method. The Matching-pursuit
algorithm is originally proposed as signal process method. This algorithm
searches a over complete basis set to find the most important ones to span a
given state.  This is basically in the same spirit with renormalization
algorithm in the sense that it keeps the most important parts and ignore small
components. Redundancy of the basis set affects the performance of the
algorithm. For sufficiently redundant basis set, the convergence of the
matching-pursuit algorithm can be exponential~\cite{33}. As a result, a few
hundreds of excitation operators are enough to arrive meaningful result for the
Hubbard model in our test calculations. This is the advantage of non-orthogonal
form of excitation operators.

The wave function of coupled cluster method (CCM) has the form
\begin{equation}\label{ccm}
 \Psi=\exp(T)\left| \Phi_0 \right\rangle,
\end{equation}
where
$\left|\Phi_0 \right\rangle=\prod_{i=1}^M a^\dagger_i|0\rangle$
is the reference state, $M$ the particle number, 
and $a^\dagger_i$ ($i=1,\cdots,N$)
the single particle creation operator. The states
$\{a^\dagger_1|0\rangle,\cdots,a^\dagger_N|0\rangle\}$ 
form single particle basis states. The operator $T$ is a 
superposition of some particle-hole excitation operators $\{ T_1,\cdots,T_n\}$
\begin{equation}\label{T}
 T=\sum_{i=1}^n\alpha_i T_i.
\end{equation}
The conventional CCM usually chooses $\{T_i\}$ as some low order excitation
operators, such as single and double excitation operators in the form
$a^\dagger_\alpha a_i$, $a^\dagger_\alpha a^\dagger_\beta a_i a_j$, $\cdots$.
Here $i$, $j\leq M$ denote the occupied orbits, and $\alpha$, $\beta>M$ denote the
unoccupied ones. A high order excitation operator is indeed a product of single
particle-hole excitation operators $a^\dagger_\alpha a_i$. Despite success for
moderate sized atoms and molecules, the number of these excitation operators
increases rapidly with particle number, and it is hard to implement in typical
systems of condensed matter physics, such as, e.g., Hubbard model away from half
filling.

We employ an efficient parameterization for the single particle-hole excitation
operators that form the excitation operators $T_i$,
\begin{equation}\label{ti}
 T_i=\prod_j^m t^{(i)}_j,
\end{equation}
where $m$ is the order of excitation. Unlike the conventional CCM, the single
particle-hole excitation operator has the form
\begin{equation}\label{tij}
t^{(i)}_j=\lambda^{(i)}_j+f^{(i)\dagger}_j g^{(i)}_j.
\end{equation}
Here $\lambda^{(i)}_j$ is an adjusting parameter usually not vanished. The
single particle creation operator 
\begin{equation}\label{fi}
f^{(i)\dagger}_j=\sum_{\alpha=M+1}^Nf^{(i)}_{j\alpha}a^\dagger_\alpha,
\end{equation}
and the  single hole creation operator
\begin{equation}\label{gi}
g^{(i)}_j=\sum_{k=1}^Mg^{(i)}_{jk}a_k
\end{equation}
are linear combination of basis single particle and hole creation operators,
respectively. The parameters $\lambda^{(i)}_j$, $f^{(i)}_{j\alpha}$, and
$g^{(i)}_{jk}$ are yet to be determined in optimization procedures. Our results
show that the parameter $\lambda^{(i)}_j$ is usually nonzero. Thus the single
particle-hole excitation operator $t^{(i)}_j$ has only finite probability to
excite a particle-hole pair. This means that the excitation operator
$T_i$ contains all orders of excitations up to order $m$. In fact, one can force
$\lambda^{(i)}_j$ to be zero. But this costs the efficiency. In this case,
one has to consider all orders of excitation explicitly. Such setting needs much
more excitation operators to reach a similar accuracy. It is easy to see from
the form of (\ref{tij}) that a low order excitation operator is a special form
of a high order one. For example, by setting $\lambda^{(i)}_m=1$ and
$f^{(i)}_{m\alpha}=g^{(i)}_{mk}=0$, a $m$-th order excitation operator becomes
in fact a $(m-1)$-th order excitation. In our calculations, we fix the order of
excitation $m$ for all operators $T_i$. Note that $f^{(i)\dagger}_j$ and
$g^{(i)}_j$ anti-commute with each other $[f^{(i)\dagger}_j,g^{(i)}_j]_+=0$.
Thus all the particle-hole excitation operators $t^{(i)}_j$ and $T_i$ commute
with each other $[t^{(i)}_j,t^{(i')}_{j'}]=0$, $[T_i,T_j]=0$.

We search the excitation operators by optimization of an approximate ground
state wave function of configuration interaction in terms of these excitation
operators. The wave function is formed by acting the excitation operators to the
reference state,
\begin{equation}\label{ci} 
\Psi_{CI}=(I+\sum_i c_i T_i)\left|\Phi_0\right\rangle.  
\end{equation} 
In this sense, our method is indeed a procedure of improvement to the CI wave
function via CCM formulation, i.e., a CCM wave function using pre-determined
excitation operator of a CI wave function.  Note that the conventional CCM wave
function also uses the same excitation operators of the CI wave function. The
CCM wave function can be viewed as a normalized form of CI wave function with
amplitudes of high order excitation operators determined by the low order ones.
There is a basic observation from the CCM theory: If two excitation operators
$T_i$ and $T_j$ have significant contribution to the CI wave function, then
their product $T_i T_j$ also has reasonable contribution to the  CI wave
function. The CI wave function (\ref{ci}) is a first order approximation to the
CCM wave function (\ref{ccm}), this linearized form is a major part of the CCM
wave function. Thus find the approximate wave function (\ref{ci}) should
determine the correct form of the excitation operators $\{T_i\}$.

We use matching-pursuit algorithm in a similar way to that of Ref.~\cite{34} to
find the operators $T_i$.  The match-pursuit algorithm searches a over complete
basis set to find a sparse representation of a given state. The convergence
depends on the redundancy of the basis set: The more redundant a basis set is,
the faster the convergence rate is~\cite{33}. The states $\{T_i|\Phi_0\rangle\}$
are not orthogonal with each other, and all possible such states form a over
complete basis set.  The non-vanish parameters $\lambda^{(i)}_j$ in the single
particle excitation operator $t^{(i)}_j$ increase the redundancy of the
excitation operators. If we set all $\lambda^{(i)}_j=0$, the redundancy decrease
drastically. In this case, the states $\{T_i|\Phi_0\rangle\}$ are almost
orthogonal with each other, and this is especially the case for high order
excitations. This is a key figure for the fast convergence with the number of
the excitation operators.  The matching-pursuit algorithm finds the basis states
one by one. Our basis set are multi-linear function of the unknown variables.
Thus if two basis states differ only by one single particle creation
(destruction) operator, they can be merged into one operators. This property can
be used to further optimize a given excitation operator.

An alternative form of the excitation operators $\{T_i\}$ in (\ref{ci}) is
\begin{equation}\label{eq5}
T'{}_i=\prod_{k=1}^{m}f'{}_{k}^{(i)\dagger}\prod_{j=1}^{m}g_{j}^{(i)}.
\end{equation}
Here the single particle creation operator 
$f'{}_{k}^{(i)\dagger}$
is a linear combination of all basis creation operators 
\begin{equation}\label{fpi}
f'{}^{(i)\dagger}_k=\sum_{s=1}^Nf'{}^{(i)}_{ks}a_{s}^{\dagger}.
\end{equation}
In fact, for every operator $T_i$, there exist an operator $T'{}_i$ such that
$T_i\left|\Phi_{0}\right\rangle = T'{}_i \left|\Phi_{0}\right\rangle$, 
and vice versa. Our calculation is indeed to search the operators
$T'{}_i$. The advantage of 
using $T'{}_i$ is that the CI wave function is multi-linear function of the
coefficients $f'{}^{(i)}_{ks}$ and $g^{(i)}_{kj}$. A rotation in the subspace
spanned by single particle creation (destruction) operators 
$f'{}_{k}^{(i)\dagger}$ ($g_{j}^{(i)}$) doesn't change the excitation operator.
The operators $T'_i$ are not commute with each other, and thus not suitable for
the CCM calculation. We transforming $T'_i$ into the form of $T_i$ for CCM
calculation.

These operators are randomly initialized. We using linear optimization
method to find these operator one by one. The detail procedures is effectively
the same as that of Ref. \cite{34}. It is a variational procedure that 
minimizes the Reyleigh quotient
$E=\left\langle \Psi_{CI}|H |\Psi_{CI} \right\rangle/ 
\left\langle \Psi_{CI} | \Psi_{CI} \right\rangle$.
The wave function (\ref{ci}) is a multi-linear function of the parameters that
define these excitation operators. Each step of search is to optimize the 
linearly dependent parameters in one particle-hole excitation operator. We
optimize each operators consecutively one by one. Backward optimization
are needed to further improve the result.

Calculation of matrix element is a basic task. Note that an excitation operator
acting on the reference state results another determinant state. And the overlap
of two determinant states is a determinant. A further detail to optimize the
computation is to employ Wick's theorem to calculate those
determinants~\cite{35}.

Using the excitation operators obtained in the above CI wave function, we
construct the CCM wave function (\ref{ccm}). Finding the amplitudes of the
excitation operators in the CCM wave function is similar to that of conventional
CCM method by a projective method. This method chooses the amplitudes of the
excitation operators such that the reference state $|\Phi_0\rangle$ to be the
ground state of the similarity transformed Hamiltonian
$\widetilde{H}=e^{-T}He^{T}$. Of course, this is equivalent to say that the
state $e^{T}|\Phi_0\rangle$ is the ground state of the original Hamiltonian $H$.
Thus the state $\widetilde{H}|\Phi_0\rangle$ is orthogonal to state
$(T_i-\langle\Phi_0|T_i|\Phi_0\rangle)|\Phi_0\rangle$,
\begin{equation}\label{pj}
 \langle\Phi_0|(T^\dagger_i-\langle\Phi_0|T_i|\Phi_0\rangle)
 e^{-T}He^{T}|\Phi_0\rangle=0.
\end{equation}
Note that the state $T_i|\Phi_0\rangle$ is not orthogonal to the reference state
$|\Phi_0\rangle$.  These equations determine the amplitudes of the excitation
operators.  This is in fact a diagonalization of the similarity transformed
Hamiltonian $\widetilde{H}$ in the subspace spanned by $|\Phi_0\rangle$ and
$\{T_i |\Phi_0\rangle\}$.  Similar to the conventional CCM, these equations are
nonlinear. The left side of each equation is an algebraic forth order polynomial
of the amplitudes. The polynomial is determined by
\begin{equation}\label{expan}
 e^{-T}He^{T}=H+[H,T]+\frac{1}{2!}[[H,T],T]+\cdots.
\end{equation}
For Hamiltonian containing one and two body operators, the above expansion
terminates at forth order.

In usual cases, several hundreds of excitation operators are enough. Thus the
number of variables in the above nonlinear equations is much smaller than the
implementation of conventional CCM.  The solution of the above equation is
easier to hand. We simply use Newton's iteration method to find solution. The
convergence is usually quite fast from a proper initial guess. From our tests,
we find that result from CI calculation is a good choice for initial guess.
Another useful rule is that the expansion to second order in Eq. (\ref{expan})
is enough to obtain accurate result. There are similar treatments in unitary
CCM, see, e.g., Refs. \cite{40,41,42}.

The main numeric cost for the amplitudes of the excitation operators is the
matrix elements of the commutators in Eq. (\ref{pj}). Each commutator of the
Hamiltonian and the excitation operators needs to be calculated independently.
Our practical implementation is to compute the matrix elements of the products
of the excitation operators and the Hamiltonian,
$\langle\Phi_0|T^\dagger_k(\prod_iT_i) H (\prod_m T_m)|\Phi_0\rangle$.
The result of the matrix element is a determinant. Application of Wick's
theorem can optimize the calculation of the matrix elements.
This is in sharp difference with the conventional CCM.
Truncation to second order in (\ref{expan}) saves much of the numeric costs.

Calculations of the expectation value of an observable need the left side
ground state $\langle\Phi'_0|$ of the similarity transformed Hamiltonian 
$\widetilde{H}$,
$\langle\Phi'_0|\widetilde{H}=\langle\Phi'_0|E_0$. Here 
$\langle\Phi'_0|=\sum_i \langle\Phi_0|(I+T_i^\dagger\phi'_i)$. We use standard
method to find the left side ground state in the subspace spanned by 
$\langle\Phi_0|$ and $\{\langle\Phi_0|T_i^\dagger\}$. The
expectation value of an observable $F$ with respect to the original
Hamiltonian's ground state is 
$\langle\Phi'_0|\exp(-T) F \exp(T) |\Phi_0\rangle$.

An advantage of the above formulation is easy to take into account of high order
excitation operators. To estimate the accuracy of the above method, consider
configuration interaction by excitation operators $T_i$, and all of their possible
products $T_iT_j$, $T_iT_jT_k$, $\cdots$, i.e., using basis states
$T_i|\Psi_0\rangle$, $T_iT_j|\Psi_0\rangle$, $T_iT_jT_k|\Psi_0\rangle$,
$\cdots$, to span the ground state of the Hamiltonian. Such configuration
interaction approaches the full configuration interaction (FCI) when basis states
$\{T_i|\Psi_0\rangle\}$ span a low order configuration interaction. Similar
to the conventional CCM, if basis states $\{T_i|\Psi_0\rangle\}$ plus the
reference state $|\Psi_0\rangle$ span major part of the ground state, our CCM
wave function approaches the configuration interaction by the excitation operators
$T_i$ and all their possible products. In other word, if CI wave function in
terms of the excitation operators $\{T_i\}$ approaches a $k$-th order
conventional CI, our CCM wave function approaches $k$-th order conventional CCM.
Such order $k$ can be significant higher than two. This is in sharp distinction
with the conventional CCM that usually restricts to the single and double
excitations, since numeric cost in conventional CCM increases drastically with
the order of excitations.





We test the above method by finding ground state of Hubbard model on
a square lattice with periodic boundary condition. The Hamiltonian reads
\begin{equation}
H  = -t\sum_{\left\langle i,j\right\rangle ,\sigma}
(a_{i\sigma}^{\dagger}a_{j\sigma}+ H.c.)+
U\sum_{i}n_{i\uparrow}n_{i\downarrow},
\end{equation}
where $a^{\dagger}_{i\sigma}$($a_{i\sigma}$) is creation(annihilation)
operator of spin $\sigma$ electron on site $i$, 
$n_{i\sigma}=a^{\dagger}_{i\sigma}a_{i\sigma}$. This model relates to many
important phenomena of correlated electrons. It is extensively explored by
many algorithms.

\begin{table}
\caption{
Ground state energies of some the $8\times 8$ systems at half filling.
The conventional CCM result is from~\cite{36}. The variational Quantum
Monte-Carlo result (VMC) is from~\cite{37}, and two Quantum Monte-Carlo results
labeled by QMC and CPMC are from~\cite{38} and~\cite{39}, respectively.
}\label{tab1}
\begin{tabular}[t]{llllll}
\hline\hline 
$U/t$ & QMC$\ $ &  CPMC     & VMC$\ \ \ $  & CCM$\ \ \ $    & PCCM    \\
\hline 
2 &  -1.214 & -1.1766 & -1.206 & -1.1192 & -1.16098125 \\ 
3 &  -1.043 &         & -1.0144 & -0.9488 & -0.991425   \\ 
4 &  -0.886 & -0.8574 & -0.8576 & -0.8247 & -0.8507625  \\ 
5 &  -0.757 &         & -0.7358 & -0.7204 & -0.737440625 \\ 
6 &  -0.657 & -0.6503 & -0.6314 & -0.6357 & -0.646553125  \\
7 &  -0.586 &         & -0.5620 & -0.5694 & -0.5736046875  \\
8 &  -0.529 & -0.5110 & -0.4926 & -0.5100 & -0.514346875  \\
9 &         &         & -0.4493 & -0.4614 & -0.465471875  \\
10&         &         & -0.4061 & -0.4217 & -0.4245375  \\
12&         &         & -0.3546 & -0.3566 & -0.360903125  \\
14&         &         & -0.3030 & -0.3107 & -0.313309375  \\
16&         &         & -0.2690 & -0.2727 & -0.2766109375  \\
18&         &         & -0.2349 & -0.2468 & -0.24736875  \\
20&         &         & -0.2184 & -0.2250 & -0.2240687  \\
\hline 
\hline 
\end{tabular}
\end{table}

Table \ref{tab1} shows the ground state energies at half filling of an $8\times8$
lattice.  We use $200$ 8-th order excitation operators to obtain the above
result labeled as PCCM. The number of the excitation operators is determined by
the convergence of the CI wave function. To compare with other results, the
conventional CCM result~\cite{36} and variational Quantum Monte-Carlo (VMC)
result~\cite{37} , as well as two QMC results labeled as QMC~\cite{38} and
CPMC~\cite{39} are also listed. Note that the CPMC result is for infinite sized
system, and it has a little bit discrepancies from the first one. We see that
our result is in the same accuracy as the conventional CCM result with some
improvement for large $U$. It is also compatible with the QMC results. The
conventional CCM method for Hubbard model works at half filling case with a
reference state that each electron occupies one site.  This kind of reference
state has translational symmetry that is a precondition of applying the
conventional CCM to the Hubbard model. Here we use the usual unrestricted
Hartree-Fock state as reference state which has no need for the precondition of
the translational symmetry. Unlike the excitation operators in the conventional
CCM, which are set by hand from consideration of the Hamiltonian's structure,
the excitation operators in our method come from search procedure with a
randomly initial guess. This implementation can apply to general systems.  The
search for the excitation operators is for configuration interaction wave
function by minimizing the expectation value of the Hamiltonian. Then we use
these operators to perform coupled cluster calculation. The result shows that
this approach works well for a reasonable reference state. In fact, in the half
filled case, the ground state energy from the mean field theory has about $90\%$
accuracy.  Our test calculations show that it is enough to expand the polynomial
$e^{-T}He^T$ of Eq.  (\ref{expan}) to the second order for the calculation of
the amplitudes of the excitation operators in the CCM wave function.

\begin{table}
\caption{
Ground state energies of some the $8\times 8$ systems at a filling $62/64$. The
QMC and VMC results are from the same references of that in table~\ref{tab1}.
}\label{tab2}
\begin{tabular}[t]{llllll}
\hline\hline 
$U/t$ & CPMC$\ $  & VMC$\ \ \ $ & HF & CI    & PCCM    \\
\hline 
2  &  -1.2071  &         & -1.0150  & -1.1761 & -1.18340625   \\
4  &  -0.89998 &         & -0.8206  & -0.8642 & -0.8873953125 \\
6  &  -0.7029  &         & -0.6201  & -0.6633 & -0.687928125   \\
8  &  -0.5705  & -0.5641 & -0.4996  & -0.5364 & -0.56278125   \\
10 &           &         & -0.4196  & -0.4533 & -0.4776859375    \\
12 &           &         & -0.3639  & -0.3946 & -0.417721875  \\
14 &           &         & -0.3229  & -0.3515 & -0.37358125   \\
16 &           &         & -0.2909  & -0.3166 & -0.33535    \\
18 &           &         & -0.2689  & -0.2909 & -0.3075171875   \\
20 &           &         & -02505   & -0.2735 & -0.289459375  \\
\hline 
\hline 
\end{tabular}
\end{table}

Table \ref{tab2} shows the ground state energies at a filling $62/64$ of a
$8\times 8$ square lattice. The QMC and VMC results are from the same references
of that in table \ref{tab1}. We also include the Hartree-Fock (HF) and
configuration interaction (CI) results of our calculation for comparison. The
conventional CCM method is hard to treat such cases of alway from half filling
due to the difficulty to find a seasonable reference state with translational
symmetry. In our implementation, the procedures are the same as above the half
filling case. The resultant accuracy and numeric cost are similar to the above
half filling case.  In fact, the performance of our method near half filling is
all the same. The reference state is still the unrestricted Hartree-Fock state.
This mean field state near half filling has a reasonable accuracy for ground
state energy, and almost correct symmetries. It is enough to use $200$ 8-th
order excitation operators to obtain table \ref{tab2}. We search the excitation
operators by optimization of the CI wave function in the same way as the half
filled case.  Similar to the conventional CCM, the result is insensitive to the
change of $U$.  When the on site repulsion $U$ increase, the accuracy of our
result is almost unchanged.  This is indeed a common property of wave function
based methods that immune from the sign problem.

\begin{table}
\caption{
Ground state energies of some the $8\times 8$ systems at various filling for
$U=8t$. The comparing QMC and VMC results are from the same sources of that in
table~\ref{tab1}.
}\label{tab3}
\begin{tabular}[t]{lllllll}
\hline\hline 
$N$ & CPMC$\ $  & VMC     & HF & CI & PCCM$^{(1)}$ & PCCM$^{(2)}$    \\
\hline 
50  &  -0.9168  &         & -0.7031  & -0.7696 & -0.8453 & -0.9167   \\
52  &  -0.8709  &         & -0.6638  & -0.7343 & -0.7943 & -0.8672   \\
54  &  -0.8200  &         & -0.6320  & -0.6973 & -0.7536 & -0.8202   \\
56  &  -0.7623  &         & -0.6462  & -0.6549 & -0.7005 & -0.7595   \\
58  &  -0.6995  &         & -0.5640  & -0.6184 & -0.6561 & -0.6991   \\
60  &  -0.6316  &         & -0.5331  & -0.5783 & -0.6094 & -0.6222   \\
62  &  -0.5705  & -0.5641 & -0.4996  & -0.5373 & -0.5630 & -0.5628   \\
64  &  -0.5110  & -0.4926 & -0.4659  & -0.4956 & -0.5158 & -0.5143   \\
\hline 
\hline 
\end{tabular}
\end{table}

Table \ref{tab3} shows ground state energies of a $8\times 8$ lattice at various
fillings with periodic boundary condition and the on site repulsion strength
$U=8t$. The comparing QMC and VMC results are from the same sources of that in
table \ref{tab1}. The Hartree-Fock (HF) and configuration interaction (CI)
results are from our calculations. Along with comparing results, we list two
results, labeled by superscripts $(1)$, and $(2)$, correspondent 8-th order and
higher order excitations, respectively. For fittings $50/64$, $52/64$, and
$54/54$, the excitation order is $20$. The excitation order is set to $58$ for
the fittings $58/64$ and $60/64$. For other cases near half filling, $8$-th
order is enough. The results for cases of near half filling are essentially
similar as above. However, in the cases of far away from half filling there are
instabilities in the projective numeric procedures. This is a well known problem
of the CCM's projective implementation~\cite{43,44,45}.  Similar to the
conventional CCM, the reference state is crucially important to the performance.
The mean field results exhibit shell effect. The half filling case corresponds
to the closed shell case in the quantum chemistry, and alway from half filling
cases correspond open shell cases. Conventional CCM for open shell cases usually
need multiple reference states that demands much higher numeric
cost~\cite{4,5,6,7,8,9,10}. For Hubbard model, the mean field results of open
shell cases are highly degenerated.  Different initial guess may results in
qualitatively different wave function. If one chooses such qualitatively wrong
reference state, the projective method may exhibit instability or even divergent
result.  In our calculations, we need high order excitation operators to obtain
meaningful result that is compatible with best results. The final resultant wave
function may qualitatively different from the reference state. This
qualitatively difference between CCM state and the reference state is likely to
cause divergence in the projective procedure. We have tested excitation
operators with order as high as the particle number (the highest possible
order). Table \ref{tab3} only shows convergent results. We need several trials
to obtain convergent result.  The mean field result of ground state energy has
poor accuracy in cases of far away from half filling. In some cases the accuracy
may be less than $80\%$. This is in sharp contrast to the $90\%$ plus accuracy
of half filling cases. On the other hand, the CI and CCM wave functions of open
shell cases have much more improvements to the mean field state than that of the
closed shell cases. The open shell CI wave function can result about $10\%$
improvement to the energy of the mean field state. And further more, the CCM
wave function with 8-th order excitation operators can further offers about
$10\%$ improvement to CI result.  The ground state energy from CI wave function
with High order excitation operators is very close to the that of lower order
ones.  However, the CCM wave function of high order excitation operators has
remarkable improvement to the ground state energy.  Another way of improving the
CCM wave function is using more lower order excitation operators. Our
calculations show that higher order excitation operators are more efficient than
that of more number of low order ones. At the same time, high order excitation
operators are more likely to occur instability in the projective procedure. A
converged result may go divergent by slightly modification of the excitation
operators. For example, if we further optimize the excitation operator with
stricter requirement of convergence in the calculation of the CI wave function,
the subsequent projective procedure to find CCM wave function may run into
divergent. The open shell cases need further investigation.



\begin{figure}
\includegraphics[angle=-90,width=\columnwidth,clip]{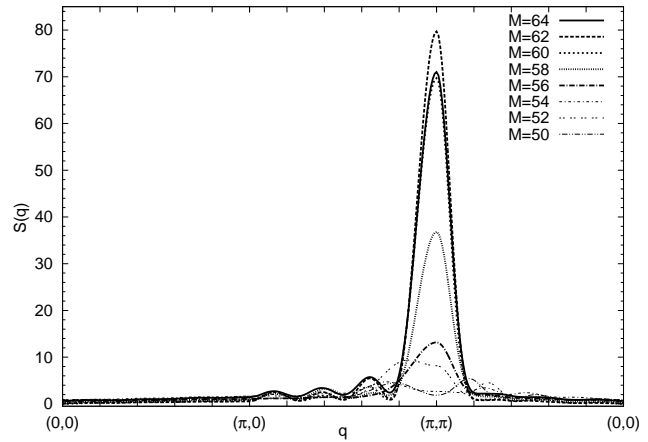}
\caption{\small
Spin structure factor versus wave vector along three different lines for various
filling numbers at $U=8t$ with a $8\times 8$ system size.
}\label{fig1} 
\end{figure}

Figure \ref{fig1} shows the magnetic structure factors
\begin{equation}
S(\mathbf{q})=\frac{1}{N}\sum_{i,j}e^{(i\mathbf{q}\cdot(\mathbf{R}_i-\mathbf{R}_j))}
\langle(n_{i\uparrow}-n_{i\downarrow})(n_{j\uparrow}-n_{j\downarrow})\rangle,
\end{equation}
along three different lines for various filling with interaction strength $U=8$
on a $8\times8$ lattice. Near half filling, the structure factors peak at
$(\pi,\pi)$ indicating anti-ferromagnetic correlation. The correspondent mean
field result also shows same correlation with a magnitude of several times
smaller. In the region far from half filling, the CCM results of magnetic
structure factors are qualitatively different from mean field result. Such
difference indicates that the mean field result is qualitatively wrong. In fact,
the CI wave functions of the open shell cases also have qualitatively different
magnetic structure factors with that of mean field results.  These cases cause
instability and need further investigations. On the other hand, comparison of
difference between the structure factor of CCM wave function and that of the
reference state provides a way to check the reference state.


\begin{figure}
\includegraphics[angle=-90,width=\columnwidth,clip]{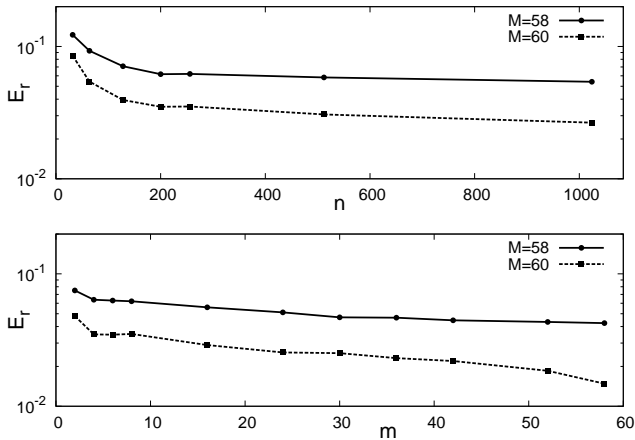}
\caption{\small
Relative error $E_r$ versus number of excitation operators $n$ (top panel)
and order of excitation operators $m$ (bottom panel), respectively,
in a $8 \times 8$ sized system. The order of excitation is $8$ in top panel,
and the number of excitation operators is $256$ in bottom panel.
}\label{fig2} 
\end{figure}

Figure \ref{fig2} shows convergence rate versus number of excitation operators
and order of excitation operators in the top and bottom panels, respectively.
We show two filling numbers, $58/64$ and $60/64$, in a periodic $8\times8$
lattice with $U=8$. Other cases are quite similar. Here the convergence rate is
relative error with respect to best result, $E_r=|(E-E_0)/E|$ with $E$ being our
numeric result and $E_0$ the best available result from others. We see that the
wave function converges quickly with the number of the excitation operators. One
usually needs about $200$ to $300$ excitation operators to arrive a convergent
wave function.  This is much faster than the CI wave functions. Note that using
high order excitation operators automatically includes lower order operators.
For a proper reference state near half filling, eighth order excitation
operators are enough to arrive a convergent wave function. The CI wave function
is rather insensitive to the order of excitation.  The  CI wave functions with
second to eight-th order excitation operators give similar ground state energy.
However, the CCM wave function with higher order excitation operators produces
much lower ground state energy. For open shell cases, one needs much higher
order of excitation operators to obtain ground state energy comparable with
other best result. A convergent CCM wave function of higher order excitation
operators can be qualitatively different from the reference state. In the
$50/64$ filling case, we use $20$-th order operators to obtain an convergent
wave function. Our results show that high order excitation operators are more
efficient to improve the CCM wave function than increasing number of lower order
excitation operators.  Note that, one can set $\alpha_i=0$ in Eq. (\ref{T}), and
consider every order of excitation operators explicitly. Such choice in our
calculations performs poorly. In our test calculations, one needs much more
excitation operators. Even in the case of half filling, one needs more than
$1000$ operators up to 4-th order to arrive convergence. This indicates that let
$\alpha_i$ as an adjustable parameter is an optimal form for the excitation
operators.


In summary, we show a coupled cluster method in terms of non-orthogonal
excitation operators. Search of these optimal excitation operators is indeed in
the same spirit of renormalization approach. In comparison with conventional
CCM, these parameterized excitation operators represent the most relevant
portion of the whole excitation operators. Several hundreds of excitation
operators are enough to arrive meaningful results. This number is several orders
smaller than that of conventional CCM. This performance makes it possible to
apply the CCM to condensed matter physics. The current implementation is indeed
a single reference CCM that is efficient in cases of near closed shell. Other
cases in deep regions of open shell need further investigations to improve the
projective procedure or developing multi-reference states CCM.



\end{document}